\documentclass[12pt]{article}
\usepackage{amsmath,amsthm,amssymb,srcltx}
\usepackage{graphicx,psfrag,epsf}
\usepackage{enumerate}
\usepackage{natbib}
\usepackage[hang]{subfigure}
\usepackage{dsfont}
\usepackage{epstopdf}

\usepackage{indentfirst}
\usepackage{overpic}
\usepackage{multirow}
\usepackage[mathscr]{euscript}
\usepackage{bm}

\newcommand{\blind}{1}

\newtheorem{thm}{Theorem}

\newtheorem{prop}{Proposition}

\newcounter{assumption}

\newcommand\e{\epsilon}

\newcommand\ols{\textnormal{ols}}
\newcommand\liu{\textnormal{liu}}
\newcommand\sign[1]{\textnormal{sign}{(#1)}}

\def\spacingset#1{\renewcommand{\baselinestretch}%
{#1}\small\normalsize} \spacingset{1}

\def\t{{ \mathrm{\scriptscriptstyle T} }}
\newcommand{\normm}[1]{{\vert\kern-0.25ex\vert #1
		\vert\kern-0.25ex\vert}}

\begin{document}



\if1\blind
{
\title{\bf Smooth Adjustment for Correlated Effects}
\author{Yuehan Yang\hspace{.2cm}\\
School of Statistics and Mathematics,
\\Central University of Finance and Economics\\
and \\
Hu Yang \\
College of Mathematics and Statistics, Chongqing University}
\maketitle
} \fi

\if0\blind
{
\bigskip
\bigskip
\bigskip
\begin{center}
{\LARGE\bf
Smooth Adjustment for Correlated Effects}
\end{center}
\medskip
} \fi

\bigskip
\begin{abstract}
This paper considers a high dimensional linear regression model with corrected variables. A variety of methods have been developed in recent years, yet it is still challenging to keep accurate estimation when there are complex correlation structures among predictors and the response. We propose an adaptive and ``reversed'' penalty for regularization to solve this problem. This penalty doesn't shrink variables but focuses on removing the shrinkage bias and encouraging grouping effect. Combining the $ l_1 $ penalty and the Minimax Concave Penalty (MCP), we propose two methods called Smooth Adjustment for Correlated Effects (SACE) and Generalized Smooth Adjustment for Correlated Effects (GSACE). Compared with the traditional adaptive estimator, the proposed methods have less influence from the initial estimator and can reduce the false negatives of the initial estimation. The proposed methods can be seen as linear functions of the new penalty's tuning parameter, and are shown to estimate the coefficients accurately in both extremely highly correlated variables situation and weakly correlated variables situation. Under mild regularity conditions we prove that the methods satisfy certain oracle property. We show by simulations and applications that the proposed methods often outperforms other methods.
\end{abstract}

\noindent%
\vfill

\newpage
\spacingset{1.45} 

\section{Introduction}
High-dimensional data analysis is a topic of great interest due to the growth of applications, i.e. portfolio allocation in finance; gene selection, etc. We focus here on the linear regression model
\[y = X \beta + \e,\]
where $ y $ is an $ n $-dimensional response vector and $ X =  (X_1,....,X_p) $ is an $ n \times p $ design regression matrix of $ p $ variables. $ \e $ is an $n$-vector of standard Gaussian random noises with mean $0$ and variance $\sigma^2$. $ \beta = (\beta_1,...,\beta_p)^\t $ is a vector of unknown regression coefficients. The sparse and high-dimensional settings mean $ n \ll p $ and many of the components of $ \beta $ are zero. Let $ q $ be the nonzero number of $ \beta $. There is always assumed that $q \leqslant n$.

For the sparse regression problem, the general approach is to determine the estimate $\hat \beta$ by solve a penalized squared loss
\[ L(\beta) = \dfrac{1}{2}\normm{y - X\beta}^2_2 + \text{pen}(\beta).\]
For the penalty $ \text{pen}(\beta) $, different choices lead to different approaches. With the choice $ \text{pen}(\beta) = \lambda \normm{\beta}_1 $, the approach is known as Lasso \citep{tibshirani1996lasso}. The naive Elastic Net approach \citep{zou2005elastic} results from choosing $ \text{pen}(\beta) = \lambda_1\normm{\beta}_1 + \lambda_2\normm{\beta}_2^2 $. Under a strong irrepresentable condition, literature proved that the Lasso and the Elastic Net are both variable selection consistent \citep{meinshausen2006high,zhao2006lasso,jia2010elastic}.
Minimax Concave Penalty (MCP) \citep{zhang2010mcp}, $ \text{pen}(\beta) = \lambda\int^{|\beta|}_0(1-\frac{x}{\gamma\lambda})_+dx$, is a typical example of nonconvex penalty which enjoys nice properties. Other variants include the SCAD of \citet{fan2001variable}, Adaptive Lasso of \citet{zou2006adaptive}, Group Lasso of \citet{yuan2006model}, Sparse-Group Lasso of \citet{simon2013sparse}, Spline-lasso of \citet{guo2016spline}, Combined L-one and Two (CLT) of \citet{ahsen2017two} and so on, many of them are proposed to solve the problems with complex correlation structure, yet it is still challenging to keep accurate estimation for the kind of data.

\begin{figure}[ht]
\centering
\subfigure[]{\label{fig1}
\includegraphics[width=.45\textwidth,height=.45\columnwidth]
{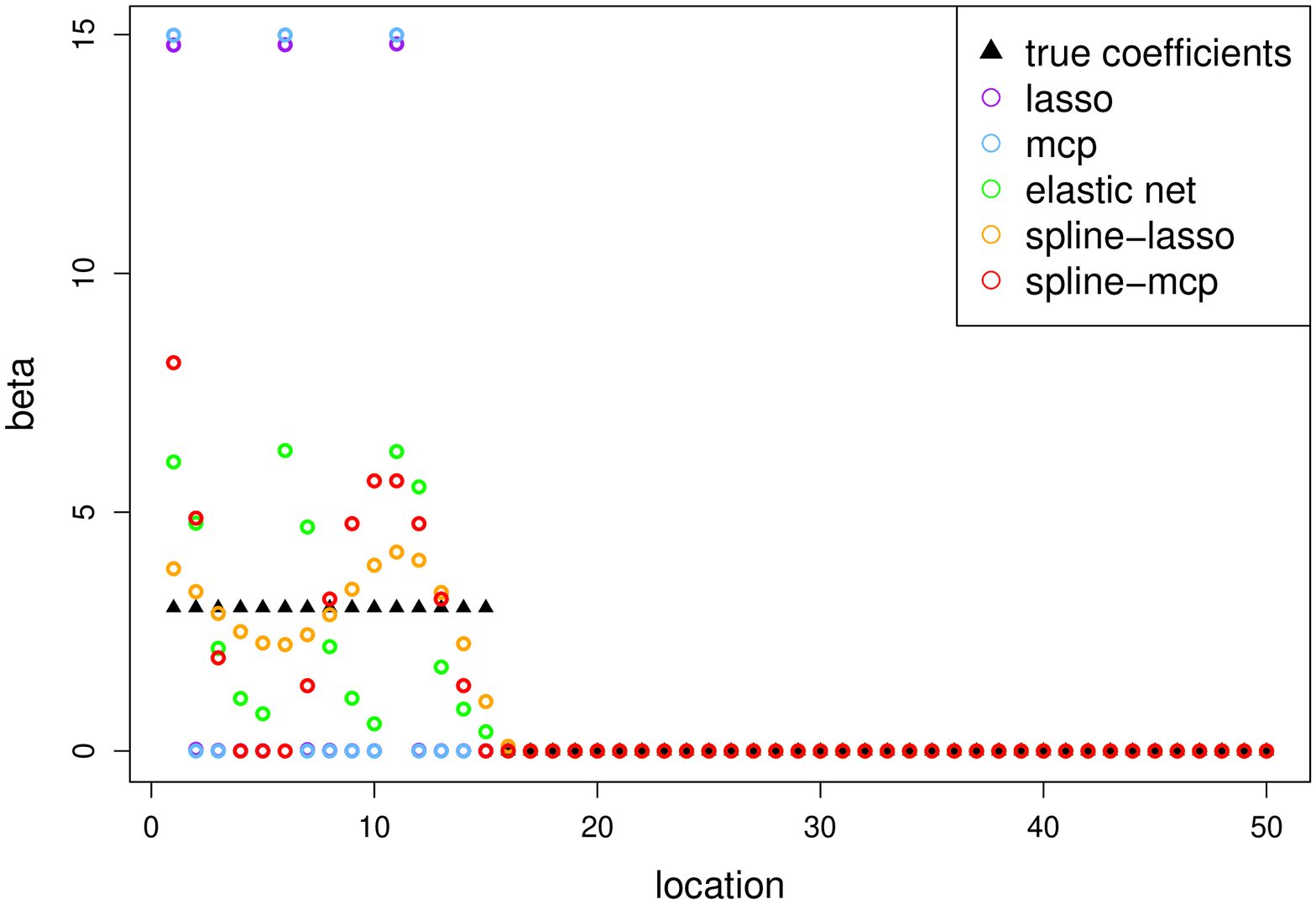}}
\subfigure[]{\label{fig2}
\includegraphics[width=.45\textwidth,height=.45\columnwidth]
{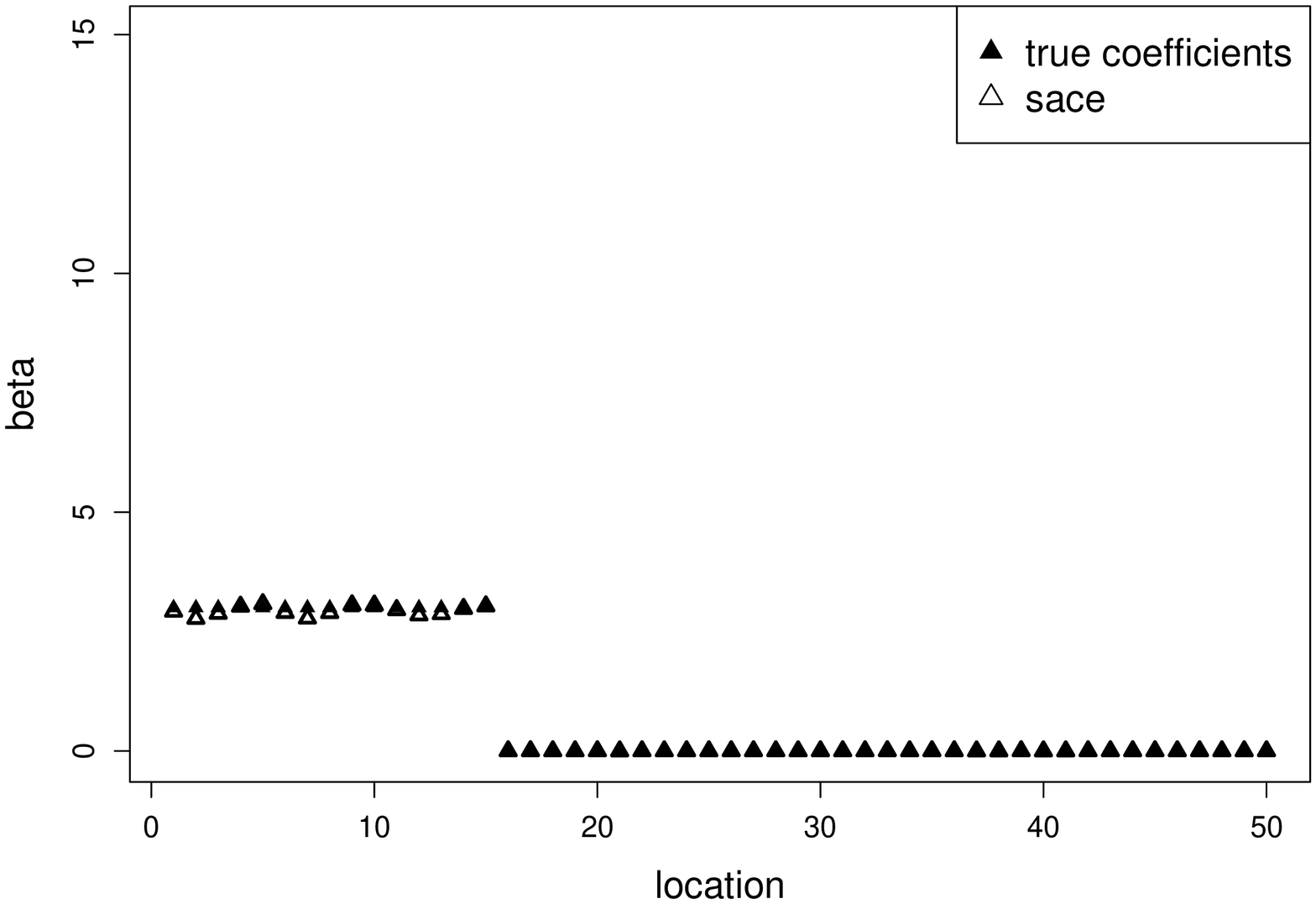}}
\caption{{Highly correlated design from Simulations. The first 1 to 5, 6 to 10 and 11 to 15 relevant variables are highly correlated (Correlation coefficients are larger than 0.9.). More details of this example can be found in Section~\ref{section:simulation}.}}
\end{figure}

For example, Spline-lasso and Spline-MCP \citep{guo2016spline} are proposed for the cases where different features within a group are different and change smoothly, however, when features are equally important in a group, the estimate still remain smooth and results in unwanted bias. Same situation happens on the other methods, i.e. Lasso and MCP tend to select only one variable from the group; Elastic Net and CLT, which encourage the grouping effect, still tend to distinguish many important variables from other ``less'' important variables in one group. We use a simple example, Figure~\ref{fig1}, to describe above situation.

As shown in Figure~\ref{fig1}, collinearity between variables adds difficulty to the problem of variable selection and estimation. We plan to address this issue by introducing a new penalty. Note that Elastic Net is a combination of the Lasso and the Ridge penalty, and there is another biased estimator for the correlated situations in low-dimensional settings, named Liu estimator \citep{liu1993new}, which combines the advantages of the Ridge estimator $\hat \beta= (X^\t X+\lambda I)^{-1}X^\t y$ and the Stein type estimator $\hat \beta= c \hat \beta^\ols$:
\[\hat \beta_\liu = (X^\t X+I)^{-1}(X^\t y + d \hat \beta_{\text{ols}}),\]
where $ d \in [0,1] $ is the tuning parameter and $ p < n $. For $d=0$, $\hat \beta_\liu$ becomes the ridge estimator; for $d=1$, $\hat \beta_\liu$ becomes the ordinary least squares. Besides, $ \hat \beta_\liu $ is a linear function of $d$, which overcome the problem of Ridge regression that it is a complicated function of $\lambda_2$.

Inspired from previous work, we proposes new penalized methods, Smooth Adjustment for Correlated Effects (SACE) and Generalized Smooth Adjustment for Correlated Effects (GSACE), to estimate the coefficients under correlated variable situation. One of the methods, SACE, is proposed as following:
\[\hat \beta := \arg \min\limits_{\beta}\bigg\{
\dfrac{1}{2}\normm{y - X\beta}^2_2 + \dfrac{1}{2}\normm{\beta}^2_2 + \lambda\normm{\beta}_1 - d (\hat\beta^0)^\t \beta
\bigg\}.\]
The new penalties includes two parts: $ \dfrac{1}{2}||\beta||^2_2 $ encourages the group effect without tuning and $ d (\hat\beta^0)^\t \beta $ is an adaptive and reversed penalty with the parameter $ d $ and an initial estimator to control and smooth the correlated effects.

A simple example of the SACE's performance is shown in Figure~\ref{fig2}; more details of this estimator can be found in Section~\ref{sacelasso}. We will first introduce the SACE estimator and its related properties; then we will introduce the GSACE, a more general version of SACE. Both estimators absorb the benefit of Liu estimator that they are linear functions of parameter $ d $, avoiding the computational waste for tuning parameter's selection. We will show that the reversed penalty can delete the noise variables and reduce the bias. Beyond that, compared with the traditional adaptive penalized methods, the proposed methods have less influence from the initial estimator $ \hat \beta^0 $ and can reduce the false negatives of the initial estimation, i.e. the SACE estimate may be nonzeros when the related initial estimate are zeros. Detailed discussions of the proposed methods will be given in the following.


The rest of the paper is organized as follows. Section 2 defines the SACE, GSACE and proves their statistical properties. Simulation results comparing the proposed methods and others are presented in Section 3. In Section 4, stock market data are used to illustrate our methodology and show the performance of the proposed methods. A summary and discussions are given in Section 5. Technical proofs of the main results can be found in the supplementary material for the paper.

\section{Methodology and Main Results}
Throughout this paper, we assume that the response and the predictors are standardized:  1) $y$ is assumed been centered at $ 0 $ to avoid the need for an intercept. 2) $X$ is assumed be standardized so $\dfrac{1}{n}X_j^\t X_j=1, \ \text{for} \ j=1,...,p $.

\subsection{SACE}\label{sacelasso}
For any fixed nonnegative $\lambda$ and $d \in [0,1]$, we define the SACE estimator
\begin{equation}\label{eq:procedure}
\hat \beta  := \arg \min\limits_{\beta}\bigg\{
\dfrac{1}{2}\normm{y - X\beta}^2_2 + \dfrac{1}{2}\normm{\beta}^2_2 + \lambda\normm{\beta}_1 - d (\hat\beta^0)^\t \beta
\bigg\},
\end{equation}
where $ \hat \beta^0 $ is the initial estimator. If we delete the last item, above procedure equals to the naive Elastic Net with regularization parameter $\lambda_2 = 1$ (More details of the naive Elastic Net can be found in \citep{zou2005elastic}). From experience, we cannot choose a much larger value for $\lambda_2$ because in that case the $l_2$ penalty would become dominant, and no estimates will be set to zero. On the other hand, we let $d$ be the tuning parameter instead of $ \lambda_2 $.

For describing the features of the penalty, $-d (\hat\beta^0)^\t \beta$, we write the solution in a explicit form:
\begin{equation}\label{eq:solution}
\hat \beta_{-\xi} = 0 \ \text{and} \  \hat \beta_{\xi} = (X^\t_\xi X_\xi + I)^{-1}(X^\t_{\xi}y + d \hat \beta^0_\xi- \lambda \tau),
\end{equation}
where $ \tau = \sign{X_{\xi}^T(y - X\hat \beta) +\hat \beta_{\xi} - d \hat \beta^0_{\xi}} $ and $ \xi $ is the equicorrelation set that $\xi = \{ i \in \{1,...,p\}: |X_i^T(y - X\hat \beta) +\hat \beta_i - d \hat \beta^0_{i}|=\lambda \}$, containing the variables which have equal absolute correlation with the residual and define the nonzero coefficient set. According to the SACE's solution, we exhibit following features for the SACE:
\begin{itemize}
\item Delete the noise variables: If $ \sign{\hat \beta_j^0} \neq \tau_j $, $ d\hat \beta_j^0 $ will cause more shrinkage of $\hat \beta_j$, $j=1,...,p$. 
\item Reduce the estimation bias: If $ \sign{\hat \beta_j^0} = \tau_j $, the bias of $ l_1 $-penalized estimator can be remarkable removed when $ \hat \beta^0_j $ is large, $j=1,...,p$.
\item Linear function of $d$: Elastic Net for instance, it is a complicated function of the parameter $\lambda_2$.  As a contrast, $\hat \beta$ is a linear function of $d$, which is easier to choose.
\end{itemize}

Further, we write the SACE estimator as an special case of the Adaptive Elastic Net \citep{zou2009adaptive} with parameters $\lambda_2=1$ and $\lambda^*$, where $\lambda^*$ is a $p$-dimensional vector. Set $\mathbf{1}$ be a $p$ by $1$ vector of $1$'s and
\[\lambda^* = \lambda  \mathbf{1} - d \hat \beta^0 \tau.\]
Note that $ \lambda^* $ includes $ \tau $ hence cannot be given advanced. We give this form for fitting the traditional adaptive estimator
\[ \hat \beta_{\xi} = (X^\t_\xi X_\xi + I)^{-1}(X^\t_{\xi}y - \lambda^* \tau). \]

Comparing with the traditional adaptive tuning parameter settings \citep{zou2006adaptive,zou2009adaptive}, i.e. $\lambda^* = \lambda |\hat \beta^0|^{-\alpha}$ where $ \alpha >0 $, our setting makes following contributions:
\begin{itemize}
\item Less influence from initial estimator: The traditional adaptive estimator highly relies on the initial estimation, which leads to larger error when the initial estimator choose wrong model. The SACE uses the initial estimator to eliminate the bias caused by the $l_1$ penalty, while the initial estimator is independent from $ \lambda $ and tuned by the other parameter $ d $, making it separated from the $ l_1 $ penalty.
\item Less false negatives: The traditional adaptive procedures yield a substantially sparser fit than using the penalized procedures only, which means they always shrink the coefficients to 0 when the corresponding components of initial estimates being zeros. We avoid this problem that the SACE estimate may be nonzeros when the related initial estimate are zeros. Example can be found in Figure~\ref{fig2} and more details can be found in simulations.
\end{itemize}


Similarly as the Elastic Net, the SACE is a Lasso-type optimization problem, which implies that it also enjoys the computational advantage of Lasso. We can solve the SACE as following.
\begin{prop}\label{prop algorithm}
Given data set $(y, X)$, initial estimate $\hat \beta^0$ and parameter $d$, define an set $ \xi^0 = \{ j: \hat \beta^0_j \neq 0, j = 1,...,p \} $ and an artificial data set $ (y^*, X^*) $ by
\[ X^*_{(n+p) \times p} = 2^{-1/2}
\begin{pmatrix} X\\I \end{pmatrix}, \ \ \ y^*_{(n+p)} =
\begin{pmatrix}y+d B \cdot \hat \beta_0 \\0\end{pmatrix}, \]
where $ B_{n \times p} $ is defined as $ B_{ij} =0 $ where $j \notin \xi^0$ and $ B_{ij} = (X^{\t}_{\xi^0})^{+}_{ij} $ where $j \in \xi^0$, $i = 1,...,n$. $ (X^{\t}_{\xi^0})^{+}$ is the (Moore-Penrose) pseudo inverse of $X^\t_{\xi^0}$, i.e. $(X^\t_{\xi^0})^{+} = X_{\xi^0}(X^\t_{\xi^0} X_{\xi^0})^{-1}$.
Then the SACE estimator can be written as
\[\hat \beta  := \arg \min\limits_{\beta}\dfrac{1}{\sqrt{2}}\times\bigg\{
\dfrac{1}{2}\normm{y^* - X^*\beta}^2_2  + \lambda\normm{\beta}_1
\bigg\}.\]
\end{prop}

Next results are concerned with the theoretical properties of SACE. For simplicity of the proof, we set $\hat \beta^0$ estimated by Lasso with the same tuning parameter $ \lambda $. We will then show that the SACE enjoys sign consistency, and show the oracle inequality for SACE which gives $l_2$ error bound on the risk.

\begin{thm}\label{thm sign}
  For any $\lambda>0$ and $ d \in [0,1] $, suppose the Lasso estimator has sign consistency, then the SACE estimator has sign consistency.
\end{thm}

\begin{thm}\label{thm sace Lasso}
Let $C =\dfrac{1}{n}X^\t X$. Assume $C$ satisfies the Restricted Eigenvalue (RE) condition: with a positive constant $\kappa$ that
\[v^TCv \geqslant \kappa\normm{v}^2_2, \]
for all $ v \in C(\mathcal{O}) $, $ C(\mathcal{O}) :=\{ v \in \mathds {R}^p: \normm{v_{\mathcal{O}^c}}_1 \leqslant 7\normm{v_{\mathcal{O}}}_1 \}$ where $ \mathcal{O} \subset \{1,...,p\} $.  Set $\frac{\lambda}{\sqrt{n}} = $$4\sigma\sqrt{\log p}$ and assume $\normm{\beta}_\infty \leqslant \lambda/4$. There exists a positive constant $K$ that, with probability at least $1 - 1/p$, for $ d \in [0,1] $, the SACE estimator satisfies the bounds
\[\normm{\hat\beta - \beta}_2 \leqslant K \sqrt{\dfrac{q\log p}{n}}.\]
\end{thm}

RE condition is widely used to bound the $l_2$-error between $\beta$ and estimate $\hat \beta$ \citep{bickel2009simultaneous,meinshausen2009lasso}. This condition requires a restriction of the generalized Gram matrix $C$ to the columns $\mathcal{O}$ is invertible. It is proved that with high probability RE condition holds for general classes of Gaussian matrices, for which the predictors may be highly dependent, and irrepresentable condition or restricted isometry condition may be violated \citep{raskutti2010restricted}.

It has been shown that the Lasso estimator achieves the similar $ l_2 $ error bound under same conditions \citep{meinshausen2009lasso,negahban2012unified}, and we apply this bound for $ \hat \beta^0 $ which helps obtain our result, more details can also be found in supplementary material.

\subsection{GSACE}
MCP is proposed by \citep{zhang2010mcp}, which showed that the MCP has a number of advantages over the Lasso. We extend the SACE to the GSACE and hope we can inherit some of the nice properties. Similar as the structure of SACE, GSACE estimator is defined as following:
\begin{equation}\label{eq sace mcp}
  \hat \beta  := \arg \min\limits_{\beta}\bigg\{
  \dfrac{1}{2}\normm{y - X\beta}^2_2 + \dfrac{1}{2}\normm{\beta}^2_2 + \sum^p_{j=1}\rho(|\beta|;\lambda,\gamma)- d (\hat\beta^0)^\t \beta
  \bigg\},
\end{equation}
where $ \hat \beta^0 $ is the initial estimator, $\gamma$ is a regularization parameter,
and the penalty function $\rho(|\beta|;\lambda,\gamma)$ can be any general quadratic penalty function. We use the MCP function in this paper, which is defined as
\[\rho(t;\lambda,\gamma)=\lambda\int^{|t|}_0(1-\dfrac{nx}{\gamma\lambda})_+dx\]
and its first-order derivative is 
\[\dot \rho (t;\lambda,\gamma)=\lambda(1-\dfrac{n|t|}{\gamma\lambda})_+,\]
which subject to the following unbiasedness and selection features:
\[\dot \rho(t;\lambda,\gamma)=0, \ \forall |t| \geqslant \dfrac{\gamma\lambda}{n}, \ \ \dot \rho(0+;\lambda,\gamma)=\lambda.\]
Note we obtain $\lambda$ with a different rate compared with \citet{zhang2010mcp}, i.e. we have $\lambda \propto \sqrt{n \log p}$ while \citet{zhang2010mcp} obtained $\lambda \propto \sqrt{\log p/n}$. It is because we minimize the residual sum of squares $\normm{y - X\beta}^2_2$ instead of $\normm{y - X\beta}^2_2/n$.


For the computationally, since GSACE can be transformed to MCP by the same technique in Proposition~\ref{prop algorithm}, it is as computationally easy as MCP.
Let $S \equiv \{j\in\{1,...,p\}: \beta_j \neq 0 \}$ be the set of indices of nonzero coefficients. Define
\[\hat \beta^\ols := \arg\min\limits_{\beta: \beta_j = 0, j \notin S } \dfrac{1}{2}\normm{y - X\beta}^2_2,\]
where $\hat \beta^\ols$ is the Ordinary Least Squares (OLS) estimator on the set $S$. Define
\[\hat \beta^* =(X_S^\t X_S + I)^{-1}(X_S^\t X_S +dI)\hat \beta^\ols,\]
where $\hat \beta^*$ is the Liu estimator on the set $S$.

It is known that OLS often does poorly in prediction. Ridge estimator achieves its better prediction performance through a bias-variance trade-off \citep{zou2005elastic}, however, has unstable performance due to the selection of $\lambda_2$. In that way, $\hat \beta^*$ is a suitable estimator for quite general classes of data.
We will show in the following that under the assumed conditions and $\hat \beta^0$ estimated by MCP with same settings for simplicity of the proof, the GSACE would enjoy sign consistency and it will be the same as the oracle estimator $\hat \beta^*$ with high probability.
\begin{thm}\label{thm sace mcp}
  Set $\lambda/\sqrt{n} = 4\sigma\sqrt{\log p}$. Suppose $\min\limits_{j \in S}|\beta_j| \geqslant \gamma\lambda/n$ and $\Lambda_{min}(\dfrac{1}{n}X^\t_SX_S) \geqslant 1/\gamma$ where $\Lambda_{min}(\cdot)$ denotes the smallest eigenvalue and $ \gamma $ is a positive constant. Then
  \[P(\sign{\hat \beta} = \sign{\beta} \ \text{or} \ \hat \beta = \hat \beta^*) \geqslant 1-1/p.\]
\end{thm}

\section{Simulations}\label{section:simulation}
In this section, we give simulations to illustrate the established results. Five other methods are compared: Lasso, MCP, Elastic Net, Spline-lasso and Spline-MCP. The purpose of simulations is to show that, SACE and GSACE not only dominate others by estimation accuracy but also are better variable selection procedures than the alternatives. R packages ``glmnet'', ``lars'' can be used to compute SACE, Lasso, Elastic Net and Spline-lasso estimators; R package ``ncvreg'' can be used to compute the GSACE and spline-MCP estimators.

We consider two examples for generating $X$: highly correlated predictors vs weakly correlated predictors. From our numerical experience, we find that the convex penalty methods encourage grouped effect while the non-convex penalty methods have better performance in which all features are uniformly correlated with each other, hence we summary the performance of SACE and of GSACE in two designs separately. During all the examples, we fix $n=50$, $p=400$ so that $p \gg n$, further, we will consider the higher dimensional example in empirical analysis. There are 15 nonzero $\beta$ which has two options: 1) nonzero coefficients are equal to 3 or 2) nonzero coefficients are valued from a uniform distribution on $[0.5,1]$.

\textbf{Example 1.} This example is extended from the Example 4 of \citet{zou2005elastic}. $X_1$,...,$X_p$ are generated as follows:
\begin{align*}
    & X_i  = Z_1 + e_i, \ \ Z_1 \sim N(0,1), \ \ i=1,...,5,    \\
    & X_i  = Z_2 + e_i, \ \ Z_2 \sim N(0,1), \ \ i=6,...,10,   \\
    & X_i  = Z_3 + e_i, \ \ Z_3 \sim N(0,1), \ \ i=11,...,15,
\end{align*}
where $e_i$ are independent identically distributed $N(0,0.01)$, $i=1,...,15$. The rest of predictors are randomly generated from the multivariate normal distribution $N(0,\Sigma)$. Two different covariance structures $\Sigma$ are considered: 1) $\Sigma=I$ and 2) $\Sigma_{ij}=0.5^{|i-j|}$.

\textbf{Example 2.} $X_1$,...,$X_p$ are generated from the multivariate normal distribution $N(0,\Sigma)$. We set the correlation between predictors to 0.1.

The response variable $y$ is hence generated from
\[y = X \beta + \epsilon,\]
where $\epsilon \sim N(0,\sigma^2)$ and we take $\sigma=0.4$ as a low noise level; $\sigma=2$ as a high noise level. The tuning parameters are selected by 10-fold cross-validation. The average of each measure is presented base on 100 simulations.

\begin{figure}[htp]
  \centering
  \subfigure[$\sigma=0.4$, $\Sigma=I$]{
    \includegraphics[width=.48\textwidth,height=.417\columnwidth]
    {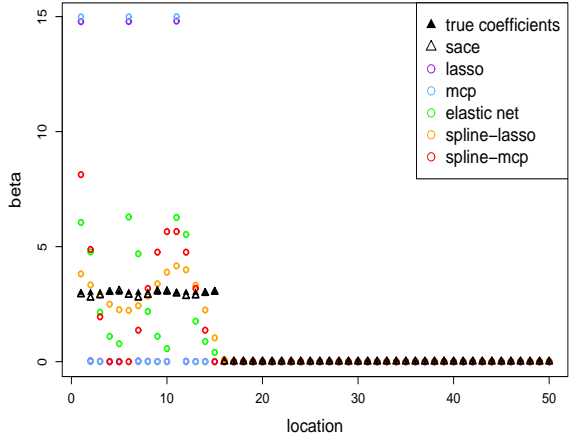}}
  \subfigure[$\sigma=0.4$, $\Sigma_{ij}=0.5^{|i-j|}$]{
    \includegraphics[width=.48\textwidth,height=.417\columnwidth]
    {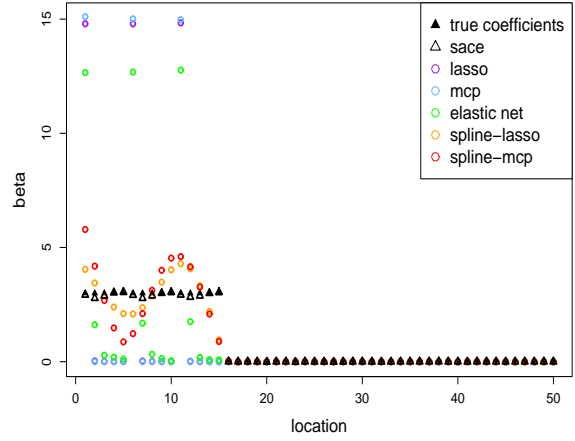}}
  \subfigure[$\sigma=2$, $\Sigma=I$]{
    \includegraphics[width=.48\textwidth,height=.417\columnwidth]
    {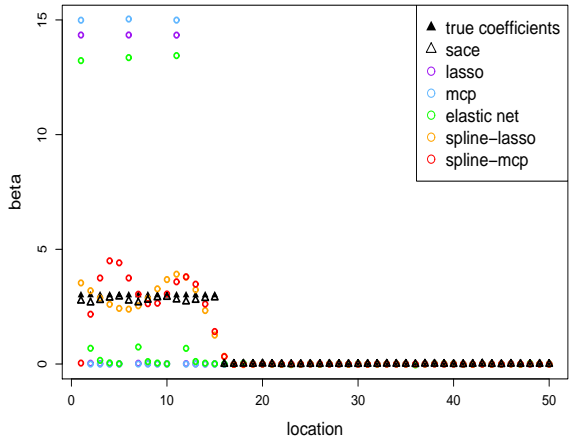}}
  \subfigure[$\sigma=2$, $\Sigma_{ij}=0.5^{|i-j|}$]{
    \includegraphics[width=.48\textwidth,height=.417\columnwidth]
    {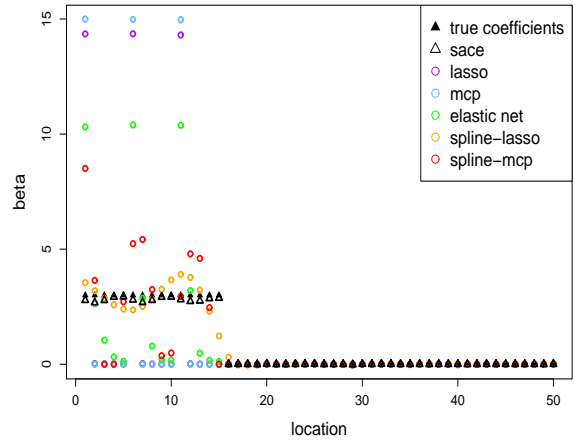}}
  \caption{{Estimation results for Example 1. The top row is with the low noise level situation ($\sigma=0.4$); the bottom row is with the high noise level situation $\sigma=2$. The figures from left to right are under different covariance structures: $\Sigma=I$ and $\Sigma_{ij}=0.5^{|i-j|}$.}}
  \label{fig:1}
\end{figure}

\begin{figure}[htp]
  \centering
  \subfigure[$\sigma=0.4$, $\beta_S=(3,..,3)$]{
    \includegraphics[width=.48\textwidth,height=.417\columnwidth]
    {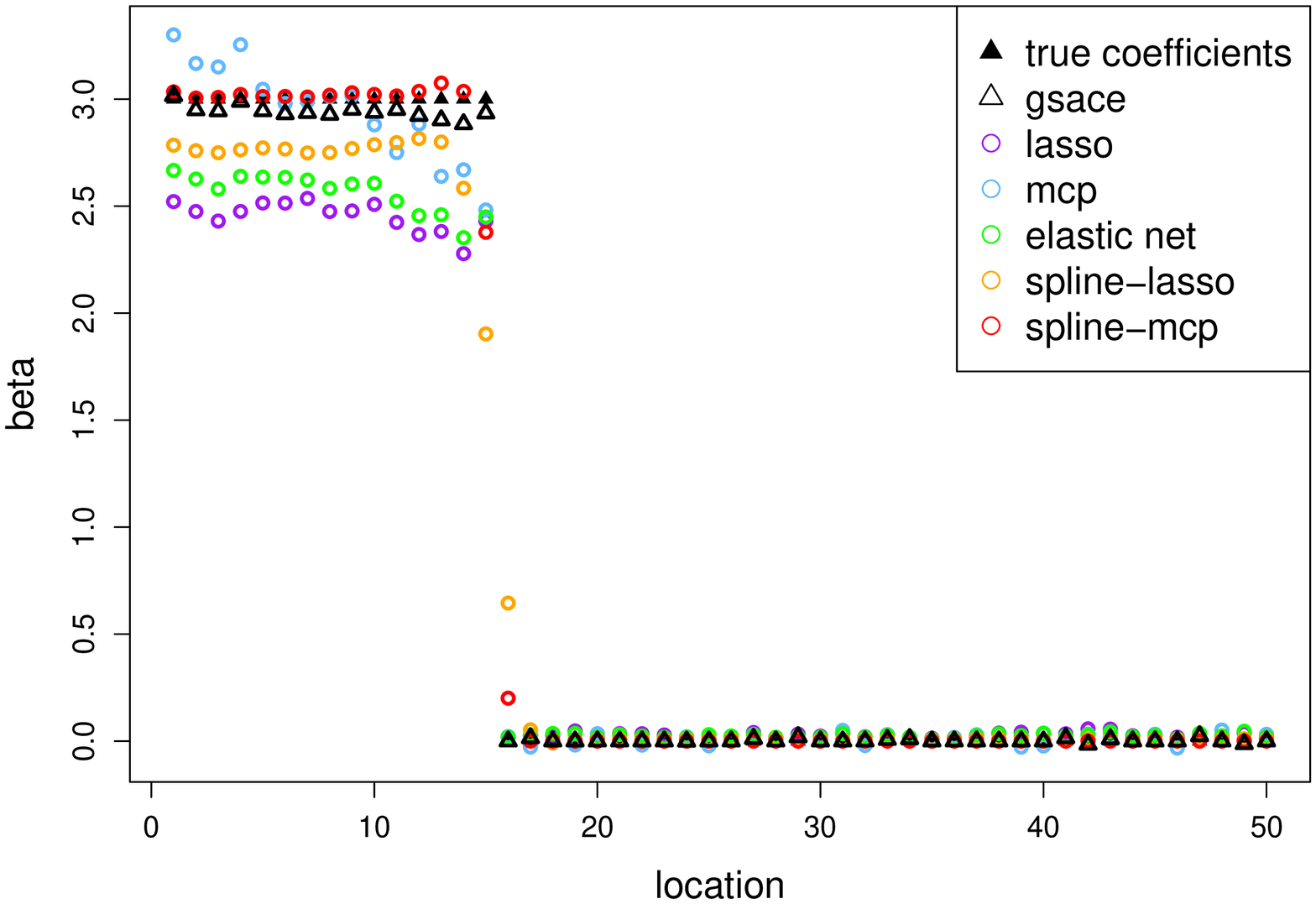}}
  \subfigure[$\sigma=2$, $\beta_S=(3,..,3)$]{
    \includegraphics[width=.48\textwidth,height=.417\columnwidth]
    {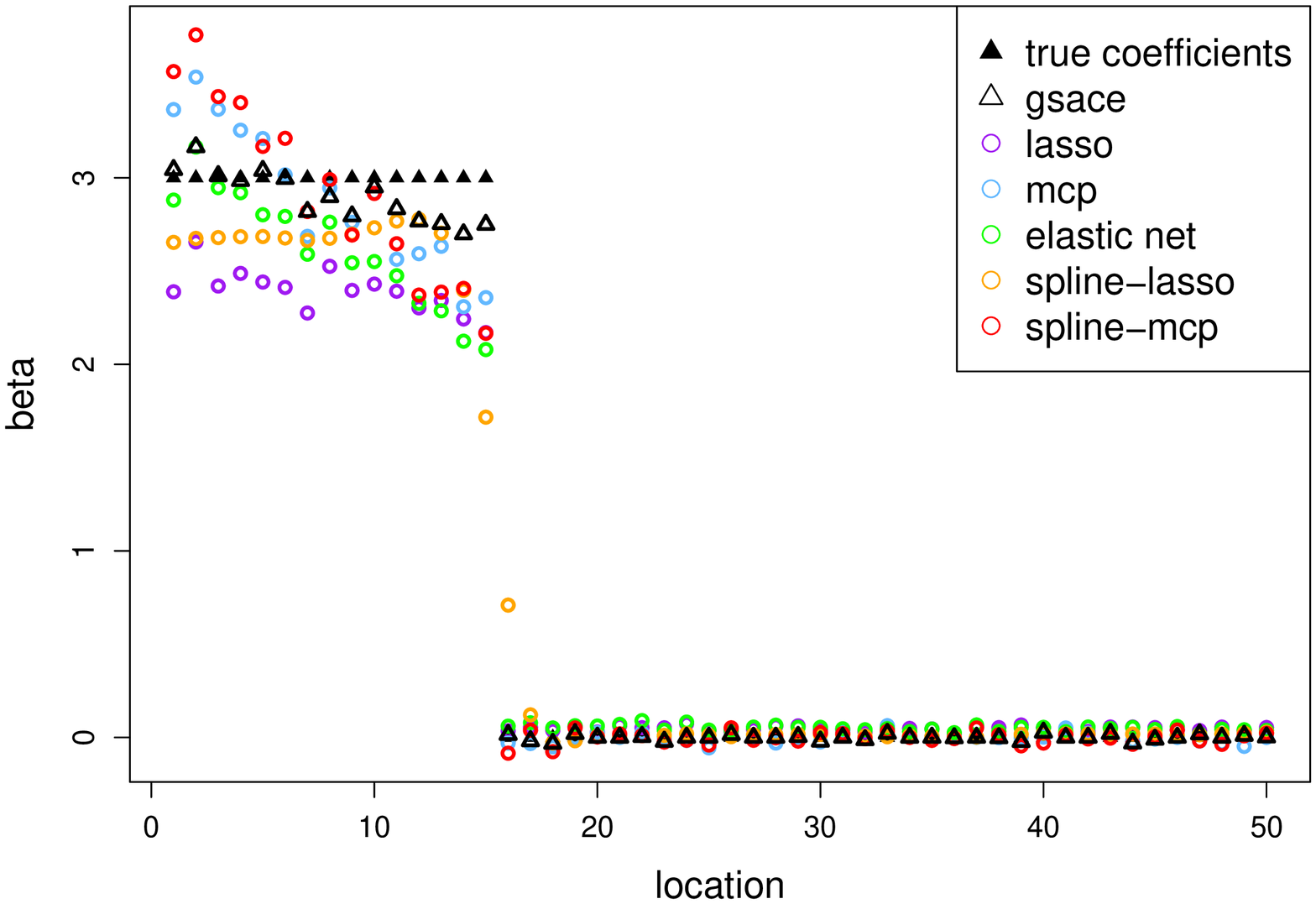}}
  \subfigure[$\sigma=0.4$, $\beta_S \sim$Unif$(0.5,1)$]{
    \includegraphics[width=.48\textwidth,height=.417\columnwidth]
    {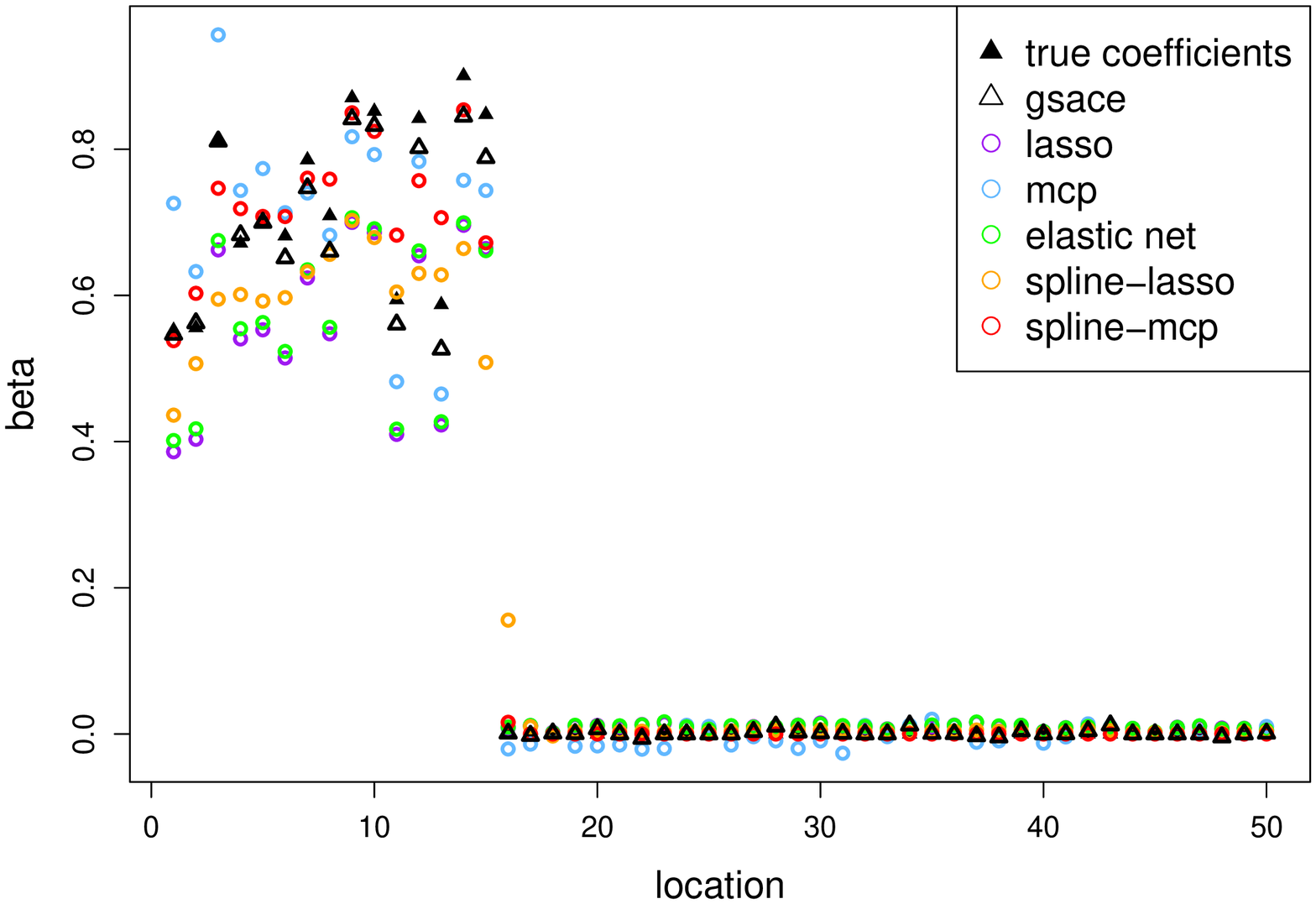}}
  \subfigure[$\sigma=2$, $\beta_S \sim$Unif$(0.5,1)$]{
    \includegraphics[width=.48\textwidth,height=.417\columnwidth]
    {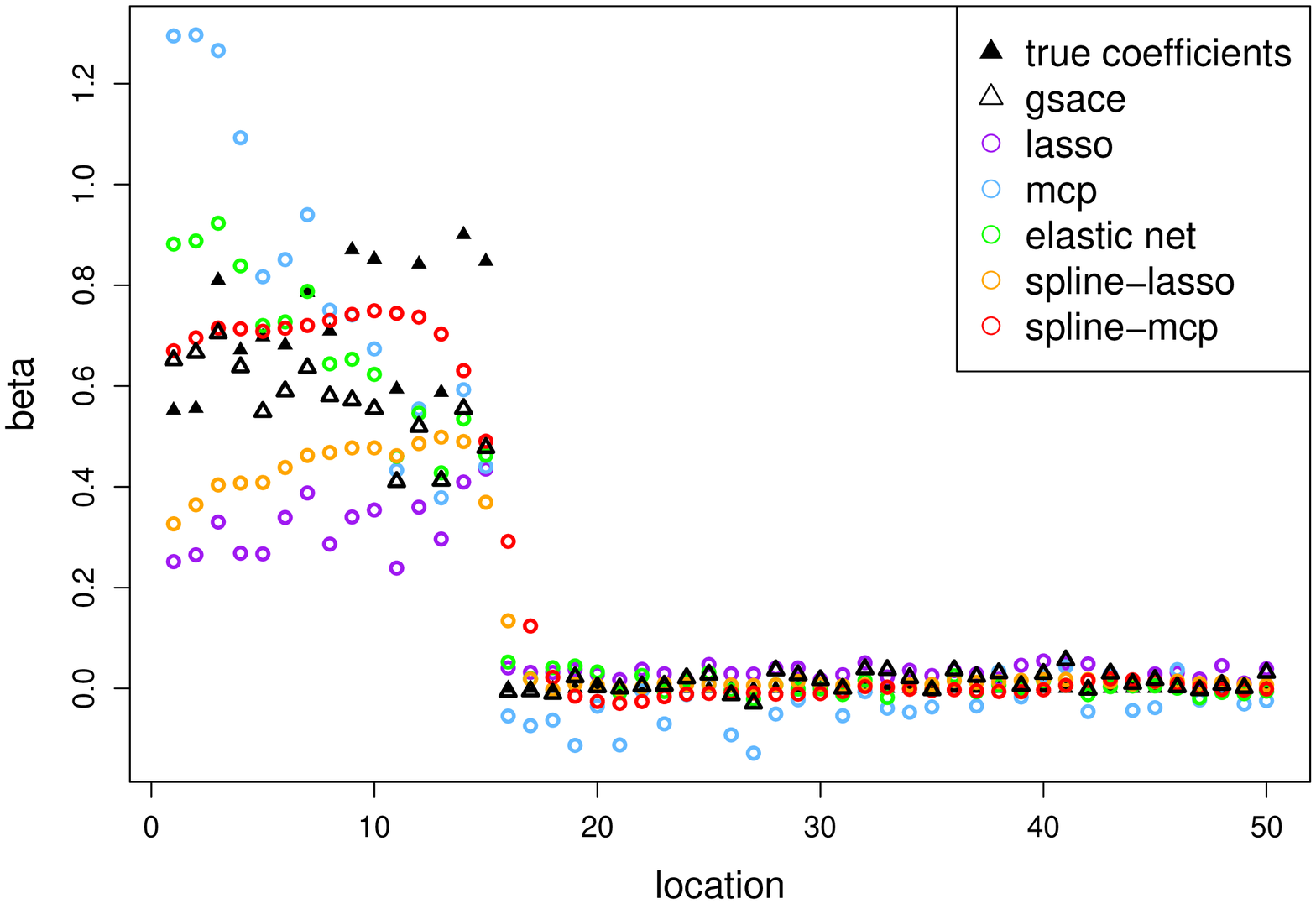}}
  \caption{Estimation results for Example 2. The top row is with $\beta = (3,...,3,0,...,0)$; the bottom row is with the situation that nonzero coefficients are valued from the uniform distribution on $[0.5,1]$. The figures from left to right are with the different noise levels: $\sigma=0.4$ and $\sigma=2$.}
  \label{fig:2}
\end{figure}

In Figure~\ref{fig:1} - \ref{fig:2}, we present the estimation results for the coefficients to demonstrate the strength of SACE and GSACE. In the highly correlated design, SACE gives the best estimation. The other methods cannot estimate the nonzero coefficients well with such high correlation between each other. In the weakly correlated design, GSACE gives the best estimation, followed by the spline-MCP which tends to provide a smoother estimation, the same as the Spline-lasso. The other three, Lasso, Elastic Net and MCP, cannot clean out the noisy signals and correlations well in both Figures.

In Table~\ref{table:1} - \ref{table:4}, we present the estimation errors, TPR and TNR for Example 1 and Example 2. Among, TPR is short for true positive rate, which is the proportion of positives being correctly identified. TNR is short for true negative rate, which is the proportion of negatives being correctly identified. EN is short for Elastic Net; S-lasso and S-MCP are short for Spline-lasso and spline-MCP respectively.

As shown in Tables, many methods show high TPR but low TNR, which means many zero coefficients are chosen. The reason for this is that the tuning parameters are determined by cross-validation, which is prediction oriented, resulting in small tuning parameters. To fix this, we apply a thresholding to the estimations. It is an effective technique in practice and the results are shown in Table~\ref{table:1} - \ref{table:4}. We do not choose the thresholding level by  cross-validation since it is less computationally efficient, instead, we use the way inspired from \citep{guo2016spline} that $\hat \beta_j = 0$ when $|\hat \beta_j|\leqslant \hat \sigma \sqrt{2 \log p}$, where $j=1,...,p$ and $\hat \sigma$ is the standard error of the estimated coefficients with small magnitude.

Table~\ref{table:1} and Table~\ref{table:3} show the estimation error of each method. We can see that the SACE and the GSACE outperform others respectively. In particular, in the highly correlated design, SACE works very well when others can't.

Table~\ref{table:2} and Table~\ref{table:4} show the selection results of each method. As shown in Table~\ref{table:2}, without thresholding, many methods have very high TPR but small TNR. When we apply the thresholding, although the TNR has been improved, the TPR of all the methods except the SACE has been significantly reduced, which means these method failed to identify the relevant variables and shrink many of them into zero. Further, it means when there's no thresholding, the estimations for the relevant variables are rather small, resulting in unwanted bias. As a contrast, with thresholding, the SACE correctly identify the relevant and irrelevant variables.

Considering Example 2 in Table~\ref{table:4}, with or without thresholding, all the methods corrected identify the relevant variables, hence we omit the TPR in the table. On the other hand, TNR has been substantially improved when the thresholding is applied.

\begin{table}[htp]
\centering
\setlength\tabcolsep{7pt}
\caption{{Estimation errors for Example 1. Case 1 - Case 4 are in different noise levels and covariance structures:
1) $\sigma=0.4$, $\Sigma=I$; 2) $\sigma=0.4$, $\Sigma_{ij}=0.5^{|i-j|}$; 3) $\sigma=2$, $\Sigma=I$; 4) $\sigma=2$, $\Sigma_{ij}=0.5^{|i-j|}$. Bottom rows give results with thresholding.}}
\label{table:1}
\begin{tabular}{ccccccc}
\hline
\hline
$\normm{\hat \beta - \beta}_2$	&	Lasso	&	MCP	&	EN	&	S-lasso	&	S-MCP	&	SACE	\\ \hline
Case 1	&	22.8970 	&	23.2248 	&	8.6856 	&	3.2178 	&	9.6247 	&	0.4464 	\\
Case 2	&	22.9196 	&	23.2819 	&	18.9794 	&	3.5655 	&	5.7227 	&	0.4455 	\\
Case 3	&	22.2007 	&	23.2473 	&	20.3641 	&	2.6616 	&	4.3643 	&	0.8170 	\\
Case 4	&	22.2012 	&	23.2111 	&	15.0182 	&	2.6936 	&	9.3840 	&	0.7945 	\\ \hline
$\normm{\hat \beta - \beta}_2$/Thresh.	&	Lasso	&	MCP	&	EN	&	S-lasso	&	S-MCP	&	SACE	\\ \hline
Case 1	&	22.9156 	&	23.2248 	&	10.0074 	&	3.2163 	&	9.6247 	&	0.4462 	\\
Case 2	&	22.9389 	&	23.2816 	&	19.7581 	&	4.1876 	&	6.4590 	&	0.4414 	\\
Case 3	&	22.2228 	&	23.2473 	&	20.7110 	&	2.6395 	&	4.3798 	&	0.8049 	\\
Case 4	&	22.2203 	&	23.2111 	&	15.6161 	&	2.6755 	&	9.6355 	&	0.7839 	\\
\hline
\end{tabular}
\end{table}

\begin{table}[htp]
\centering
\setlength\tabcolsep{10pt}
\caption{{TPR and TNR for Example 1. Case 1 - Case 4 follow the same settings as in Table \ref{table:1}. Bottom rows give results with thresholding.}}
\label{table:2}
\begin{tabular}{ccccccc}
\hline
\hline
TPR	&	Lasso	&	MCP	&	EN	&	S-lasso	&	S-MCP	&	SACE	\\ \hline
Case 1	&	1.00 	&	0.20 	&	1.00 	&	1.00 	&	0.73 	&	1.00 	\\
Case 2	&	1.00 	&	0.20 	&	1.00 	&	1.00 	&	1.00 	&	1.00 	\\
Case 3	&	1.00 	&	0.20 	&	1.00 	&	1.00 	&	1.00 	&	1.00 	\\
Case 4	&	1.00 	&	0.20 	&	1.00 	&	1.00 	&	0.80 	&	1.00 	\\ \hline
TNR	&	Lasso	&	MCP	&	EN	&	S-lasso	&	S-MCP	&	SACE	\\ \hline
Case 1	&	0.00 	&	0.99 	&	0.00 	&	0.33 	&	1.00 	&	0.15 	\\
Case 2	&	0.01 	&	0.28 	&	0.00 	&	0.13 	&	0.70 	&	0.03 	\\
Case 3	&	0.00 	&	1.00 	&	0.00 	&	0.00 	&	0.01 	&	0.02 	\\
Case 4	&	0.00 	&	0.98 	&	0.00 	&	0.00 	&	1.00 	&	0.00 	\\ \hline
TPR/Thresh.	&	Lasso	&	MCP	&	EN	&	S-lasso	&	S-MCP	&	SACE	\\ \hline
Case 1	&	0.20 	&	0.20 	&	0.60 	&	1.00 	&	0.73 	&	1.00 	\\
Case 2	&	0.20 	&	0.20 	&	0.20 	&	0.93 	&	0.87 	&	1.00 	\\
Case 3	&	0.20 	&	0.20 	&	0.20 	&	1.00 	&	0.93 	&	1.00 	\\
Case 4	&	0.20 	&	0.20 	&	0.40 	&	1.00 	&	0.67 	&	1.00 	\\ \hline
TNR/Thresh.	&	Lasso	&	MCP	&	EN	&	S-lasso	&	S-MCP	&	SACE	\\ \hline
Case 1	&	1.00 	&	1.00 	&	1.00 	&	1.00 	&	1.00 	&	1.00 	\\
Case 2	&	1.00 	&	1.00 	&	1.00 	&	1.00 	&	1.00 	&	1.00 	\\
Case 3	&	1.00 	&	1.00 	&	1.00 	&	1.00 	&	1.00 	&	1.00 	\\
Case 4	&	1.00 	&	1.00 	&	1.00 	&	1.00 	&	1.00 	&	1.00 	\\
\hline
\end{tabular}
\end{table}

\begin{table}[htp]
\centering
\renewcommand\arraystretch{0.95}
\setlength\tabcolsep{7pt}
\caption{{Estimation errors for Example 2. Case 1 - Case 4 are in different noise levels and covariance structures:
1) $\sigma=0.4$, $\beta_S=(3,..,3)$; 2) $\sigma=0.4$, $\beta (1) \sim$Unif$(0.5,1)$; 3) $\sigma=2$, $\beta_S=(3,..,3)$; 4) $\sigma=2$, $\beta (1) \sim$Unif$(0.5,1)$. Bottom rows give results with thresholding.}}
\label{table:3}
\begin{tabular}{ccccccc}
\hline
\hline
$\normm{\hat \beta - \beta}_2$	&	Lasso	&	MCP	&	EN	&	S-lasso	&	S-MCP	&	GSACE	\\ \hline
Case 1	&	2.1643 	&	0.9359 	&	1.7559 	&	1.5781 	&	0.6632 	&	0.2676 	\\
Case 2	&	2.4457 	&	1.5310 	&	1.9431 	&	1.9491 	&	1.8920 	&	0.6710 	\\
Case 3	&	0.6587 	&	0.3917 	&	0.6279 	&	0.6468 	&	0.2761 	&	0.1453 	\\
Case 4	&	1.6467 	&	1.4697 	&	0.8955 	&	1.2423 	&	0.6774 	&	0.8759 	\\ \hline
$\normm{\hat \beta - \beta}_2$/Thresh.	&	Lasso	&	MCP	&	EN	&	S-lasso	&	S-MCP	&	GSACE	\\ \hline
Case 1	&	2.1316 	&	0.8980 	&	1.7298 	&	1.4305 	&	0.6321 	&	0.2542 	\\
Case 2	&	2.3955 	&	1.4887 	&	1.9022 	&	1.7925 	&	1.8473 	&	0.6443 	\\
Case 3	&	0.6476 	&	0.3753 	&	0.6170 	&	0.6224 	&	0.2757 	&	0.1388 	\\
Case 4	&	1.6096 	&	1.4196 	&	0.8825 	&	1.2132 	&	0.6400 	&	0.8400 	\\
\hline
\end{tabular}
\end{table}

\begin{table}[htp]
\centering
\renewcommand\arraystretch{0.95}
\setlength\tabcolsep{10pt}
\caption{{TNR for Example 2. Case 1 - Case 4 follow the same settings as in Table \ref{table:3}. Bottom rows give results with thresholding.}}
\label{table:4}
\begin{tabular}{ccccccc}
\hline
\hline
TNR	&	Lasso	&	MCP	&	EN	&	S-lasso	&	S-MCP	&	GSACE	\\ \hline
Case 1	&	0.00 	&	0.65 	&	0.00 	&	0.00 	&	1.00 	&	0.87 	\\
Case 2	&	0.00 	&	0.46 	&	0.00 	&	0.00 	&	0.21 	&	0.54 	\\
Case 3	&	0.00 	&	0.30 	&	0.00 	&	0.00 	&	0.99 	&	0.51 	\\
Case 4	&	0.00 	&	0.19 	&	0.00 	&	0.00 	&	0.00 	&	0.07 	\\ \hline
TNR/Thresh.	&	Lasso	&	MCP	&	EN	&	S-lasso	&	S-MCP	&	GSACE	\\ \hline
Case 1	&	1.00 	&	1.00 	&	1.00 	&	1.00 	&	1.00 	&	1.00 	\\
Case 2	&	1.00 	&	1.00 	&	1.00 	&	1.00 	&	1.00 	&	1.00 	\\
Case 3	&	1.00 	&	1.00 	&	1.00 	&	1.00 	&	1.00 	&	1.00 	\\
Case 4	&	1.00 	&	1.00 	&	1.00 	&	1.00 	&	1.00 	&	1.00 	\\
\hline
\end{tabular}
\end{table}



\section{Real Data Example: S$\&$P500}
We apply the proposed methods to modeling the S$\&$P500 index and its constituent stocks data. First, a brief introduction of index tracking is provided: One of the popular investment products in the financial market is a collective investment scheme, called index tracking, which aims to replicate the movements of a target index, i.e. the FTSE-100 in London, S$\&$P500 in New York and CSI 300 in China. We use the proposed method for tracking the S$\&$P500 due to three reasons:
\begin{itemize}
  \item The statistical model built for the S$\&$P500 and its constituent stocks is a typical high dimensional model. S$\&$P500 includes hundreds of constituent stocks but the samples size are often less than one hundred due to the time availability.
  \item For the cost concern, the optimal replication should be holding fewer securities composed the index. It means that the investment portfolio should be a small subset for saving cost. The proposed methods are successful approaches to recover sparse solutions.
  \item There are several groups of stocks among which the pairwise correlations are very high. It requires the statistical methods encouraging the group effect. Lasso for instance, tends to select only one stock from a group and does not care which one is selected \citep{zou2005elastic}, hence is not a satisfactory variable selection method in the grouped variables situation. The ideal selection method should be able to select the proper subset including the ``representative'' securities composed the index.
\end{itemize}

Our data set consists of the prices of stocks in S$\&$P500, from Jan. 2014 to Oct. 2018 (the data come from TXDB). We divide the data set by time window, 5 months’ data ($n=100$) for modeling and one month’s data ($n=20$) for forecasting, which produces 53 forecasting samples. Let $x_{i,t}$ represent the price of $i$th constituent stock, $i=1,...,500$ and $y_t$ represent the price of the index. We describe the relationship between $x_{i,t}$ and $y_t$ by a linear regression model and tune the regularization parameters $\lambda$ and $d$ to control the amount of regularization. The target amount of subset of stocks is 50. Hence we first select a very large $\lambda$ and a very small $\lambda$ to calculate the estimators, then adapt a strategy to find one $\lambda$ that for each method can select 50 constituent stocks.

\begin{figure}[htp]
  \centering
  \includegraphics[width=\textwidth,height=.28\columnwidth]
  {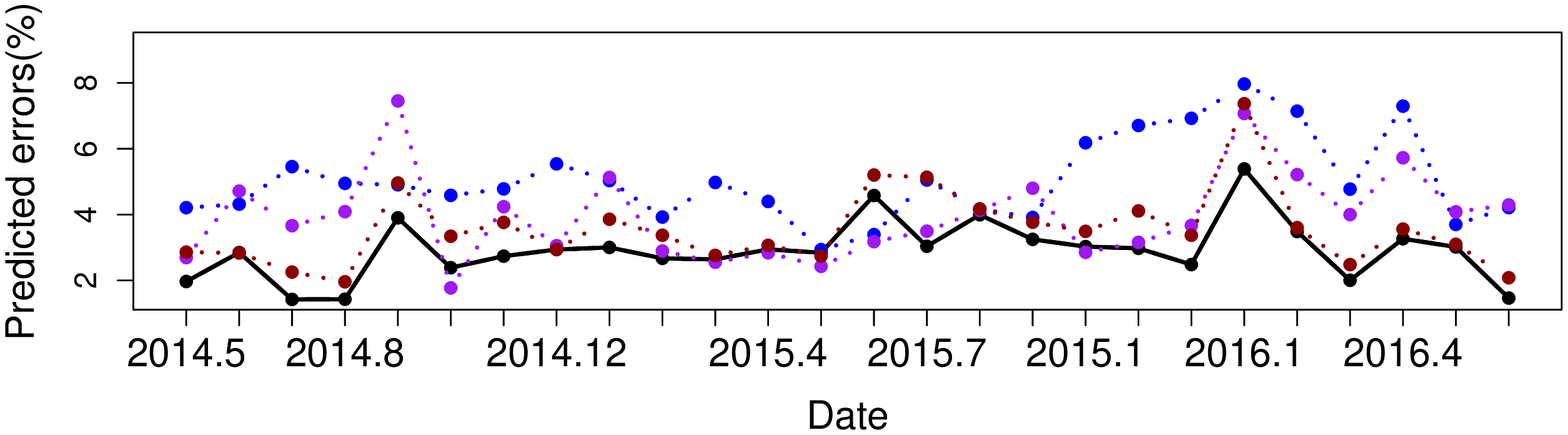}
  \includegraphics[width=\textwidth,height=.28\columnwidth]
  {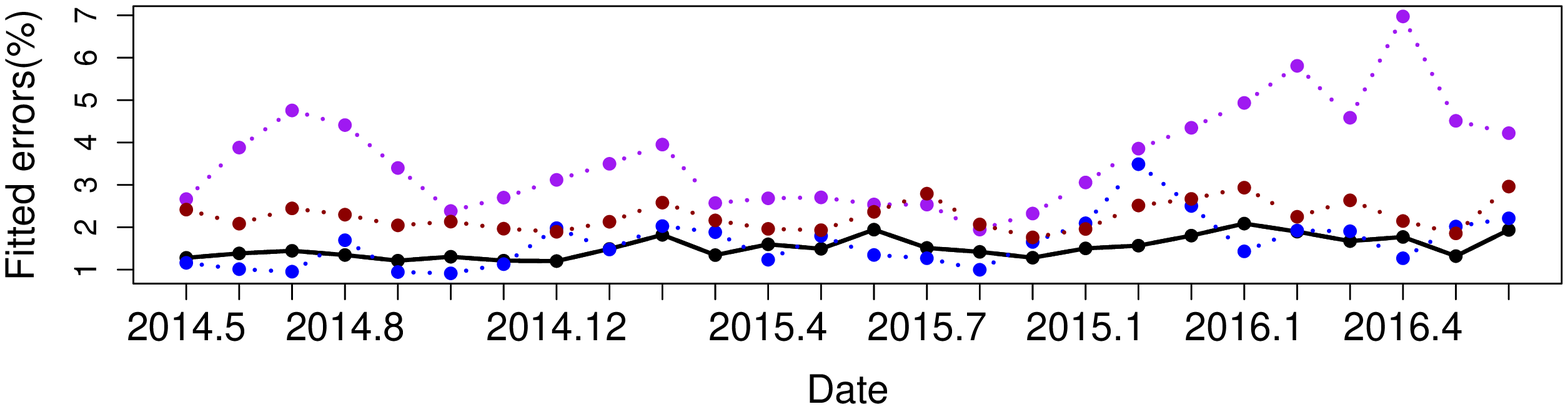}
  \includegraphics[width=\textwidth,height=.28\columnwidth]
  {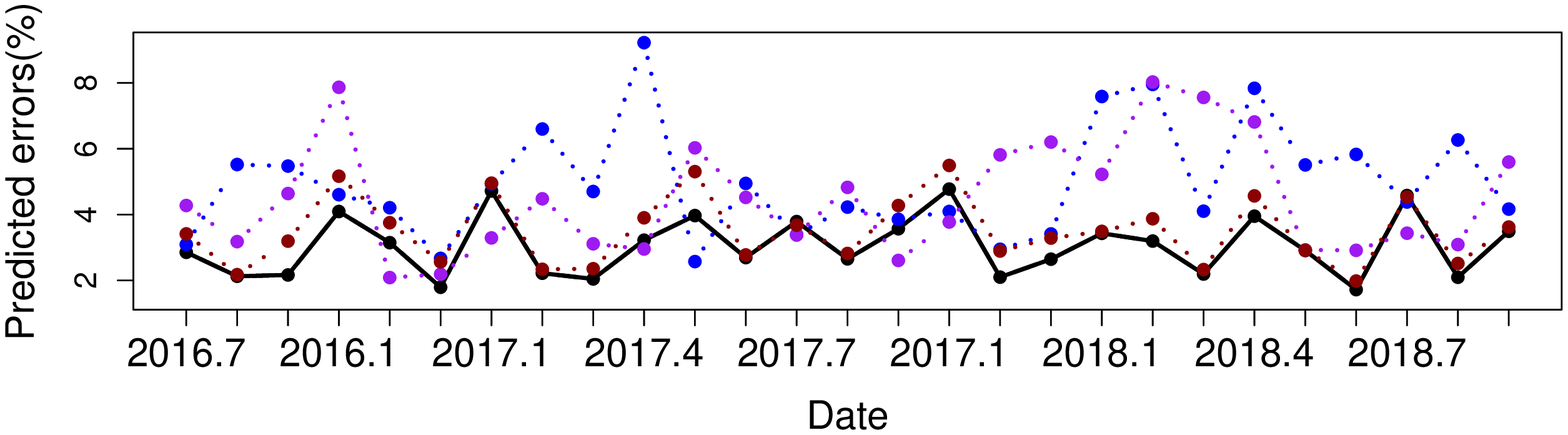}
  \includegraphics[width=\textwidth,height=.28\columnwidth]
  {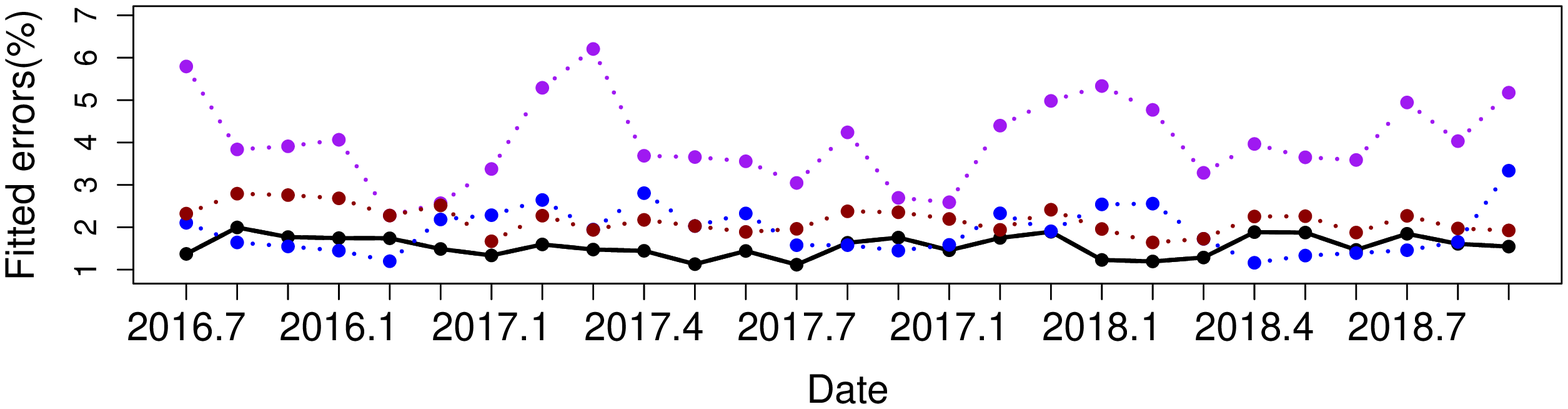}
  \caption{{Predicted tracking errors and Fitted tracking errors for four methods: the black line stands for the SACE; the blue line stands for the MCP; the purple line stands for the Elastic Net and the red line stands for the Spline-lasso.} }
  \label{fig:3}
\end{figure}

\begin{figure}[htp]
  \centering
  \includegraphics[width=\textwidth,height=.28\columnwidth]
  {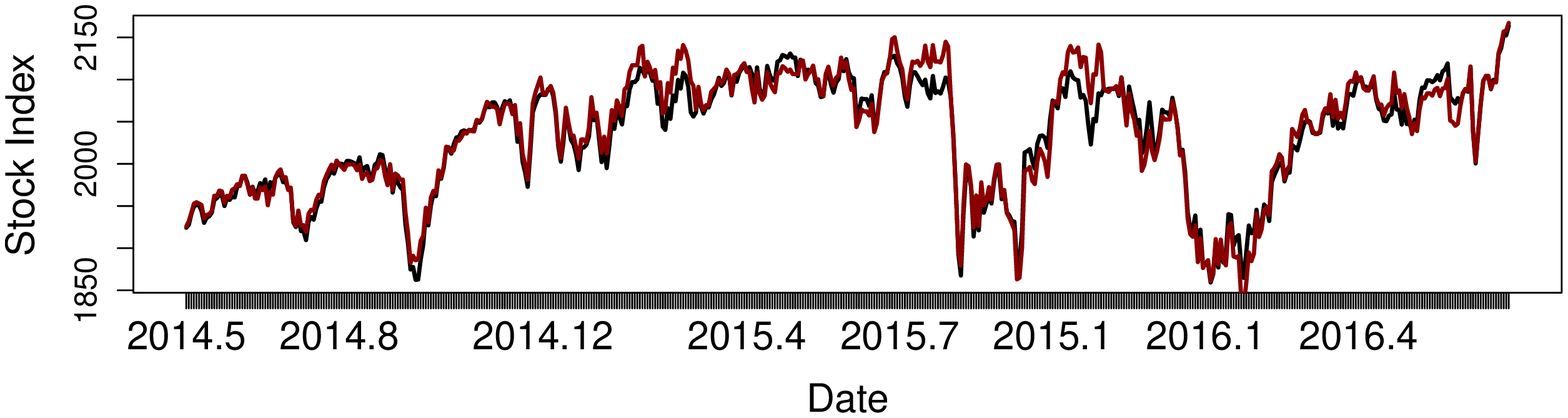}
  \includegraphics[width=\textwidth,height=.28\columnwidth]
  {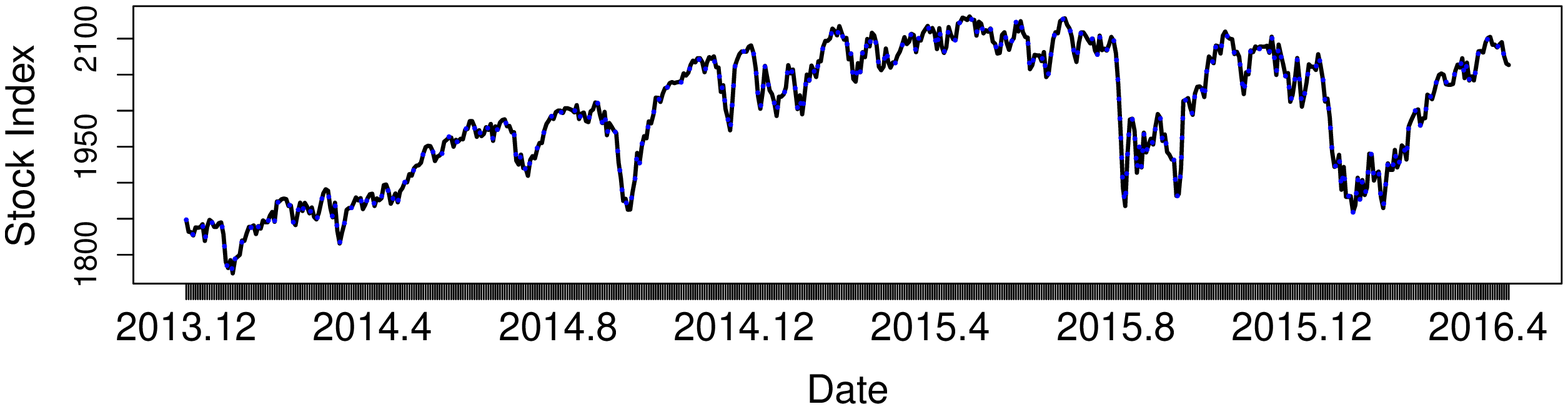}
  \includegraphics[width=\textwidth,height=.28\columnwidth]
  {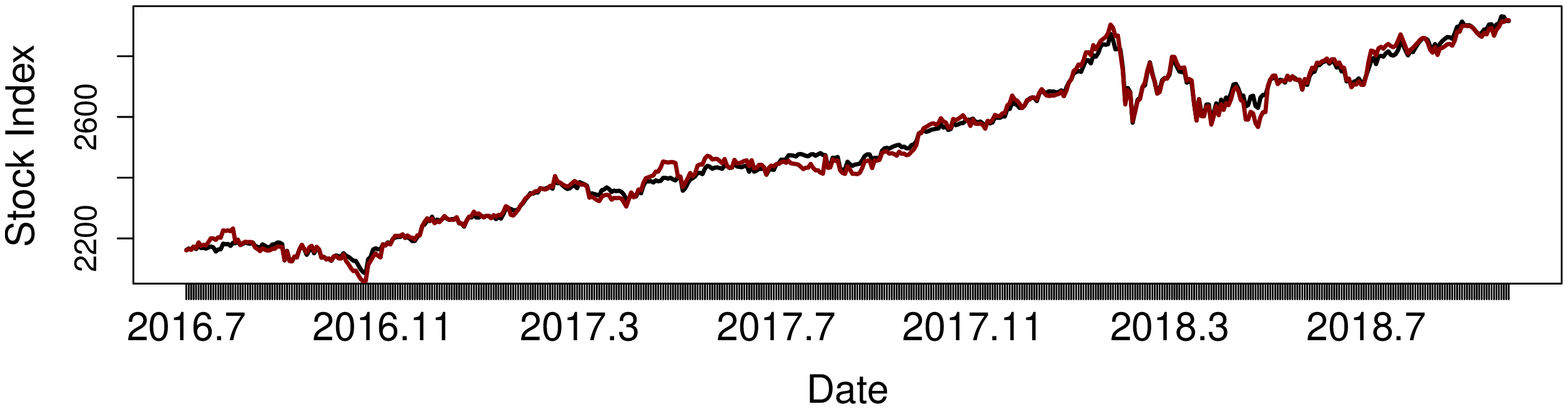}
  \includegraphics[width=\textwidth,height=.28\columnwidth]
  {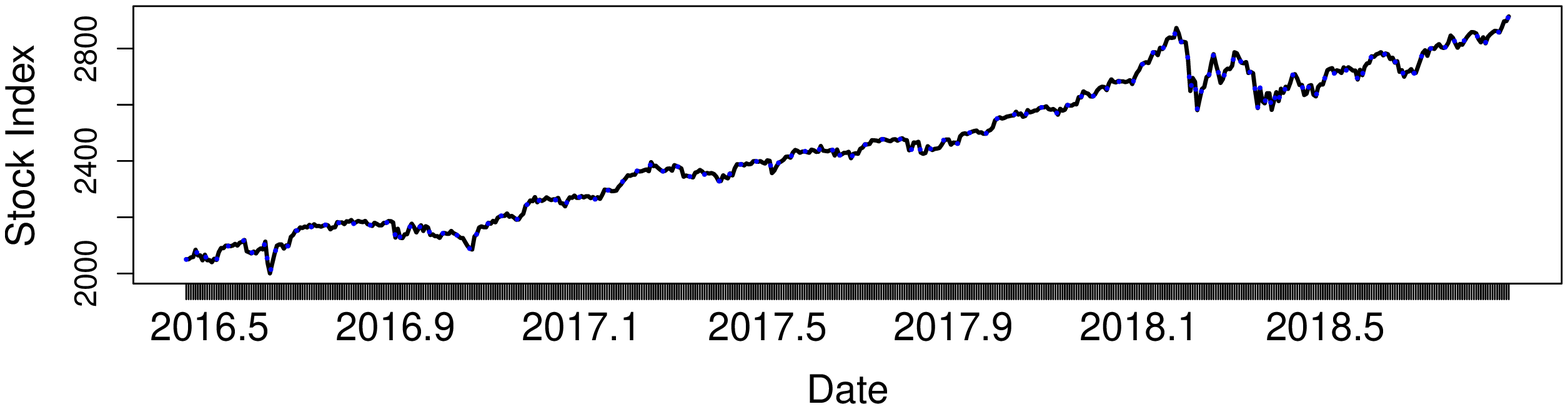}
  \caption{{Predicted/Fitted results of using SACE to track the index. The black line is the S$\&$P500 Price during recent 5 years. The red line shows the predicted tracking results and the blue line shows the fitted tracking results.} }
  \label{fig:4}
\end{figure}

We use two measures: 1) Tracking Error, results are shown in Figure~\ref{fig:3} and 2) Tracking Price $\hat y$, shown in Figure~\ref{fig:4}. The Tracking Error is defined by
\[\sqrt{250} \times \sqrt{\dfrac{\sum (\text{err}_t-\text{mean(err)})^2}{T-1}}\]
where $\text{err}_t = y_t - \hat y_t$ and $t=1,...,n$. Four methods are compared in Figure~\ref{fig:3}: SACE, MCP, Elastic Net and Spline-lasso. As one can see, the SACE nearly outperforms other estimators in predicted tracking errors during 5 years. The predicted errors of SACE are between $1\%$-$4\%$ and the fitted tracking errors are between $1\%$-$2\%$. These are qualified as a index fund in the market while the amount of stocks in our portfolio is much smaller than many index funds.
Figure~\ref{fig:4} shows the results of SACE tracking S$\&$P500. It is clear that we can use one-tenth constituent stocks (greatly reduce the transaction costs), obtained by statistical modeling, to fit/predict the target index well.

\section{Discussions}
In this paper, we consider a high-dimensional linear regression problem and there exists complex correlation structures among predictors. We propose two methods, called SACE and GSACE, which combines the Lasso or MCP penalty and the proposed penalty. We show that the new adaptive penalty can delete the noise variables and reduce the bias. Beyond that, compared with the traditional adaptive penalized methods, the proposed methods have less influence from the initial estimator $ \hat \beta^0 $ and reduce the false negatives of the initial estimation. With mild conditions, both methods enjoy sign consistency. With high probability the SACE satisfies the $ l_2 $ error bound while the GSACE equal to the oracle estimator.

The proposed methods can handle both extremely highly correlated variables setting and weakly correlated variables setting, estimating coefficients precisely. Both two data settings are tested in Simulations. Besides, the proposed estimator is a linear function of the new tuning parameter, making it easier to be chosen than the traditional tuning parameter of $l_2$ penalty. We apply the proposed methods and others to financial data, where the stocks (predictors) are always correlated. The proposed methods are successful in assets selection and produce more stable and lower rates of fitted/predicted errors.

\section*{Acknowledgements}
This work was supported by the National Natural Science Foundation of China (Grant No. 11671059).

\bibliographystyle{plainnat}
\bibliography{reference}
\end{document}